\input harvmac
 \def\12{{1\over2}}

\Title{\vbox{\baselineskip12pt\hbox{IFUG-99-013}\hbox{IFUNAM-FT-99-008}}}
 {\vbox{\centerline{The Linear Sigma Model at Work:}
	\vskip2pt\centerline{Succesful Postdictions for Pion Scattering}}}

\centerline{\bf J.L.~Lucio~M.$^{(1)}$, M.~Napsuciale$^{(1)}$
 and M.~Ruiz--Altaba$^{(2)}$}
\footnote{$^*$}
{(lucio@ifug4.ugto.mx, \ mauro@ifug4.ugto.mx, \ marti@ft.ifisicacu.unam.mx )}
\bigskip\centerline{$^{(1)}$Instituto de F\'\i sica, Universidad de Guanajuato}
\centerline{Loma del Bosque 103, 37160 Le\'on, Guanajuato, M\'exico}
 \bigskip
 \centerline{$^{(2)}$Instituto de F\`\i sica, Universidad Nacional Aut\'onoma de M\'exico}
\centerline{A.P. 20-364, 01000 M\'exico, D.F., M\'exico }

\vskip .3in

We enforce chiral Ward identities on the vertices of the linear $\sigma$ model to take into account the width of the scalar $\sigma$, considered as a true physical resonance. We consider pion scattering at very low energies and, from a fit to the measured phase shifts in the various angular momentum and isospin channels, conclude a mass  for  the $\sigma$ of  $  600{+ 200\atop - 100} $MeV. Its width turns out to be    of the same size.

\Date{17/03/99} 

\lref\Tornqvism{N.A.~Tornqvist, ``Spontaneous generation of pseudoscalar mass in the $U(3) \times U(3)$ linear $\sigma$ model,"
Phys. Lett. {\bf B426}, 105 (1998) {\tt hep-ph/9712479}.}

\lref\Napsuc{M.~Napsuciale, ``Scalar meson masses and mixing angle in a $U(3) \times U(3)$ linear $\sigma$ model," {\tt hep-ph/9803396}.}

\lref\Isck{M.~Ishida, S.~Ishida and T.~Ishida,
``Relation between scattering and production amplitudes: concerning $\sigma$ particle in $\pi \pi$ system,"
Prog. Theor. Phys. {\bf 99}, 1031 (1998)
{\tt hep-ph/9805319}.}

\lref\Lucio{J.L.~Lucio, G.~Moreno, M.~Napsuciale and J.J.~Toscano,
``Threshold behavior of $\gamma \gamma \to \pi^0 \pi^0$ as a test for the existence of a light $\sigma$ meson,"
{\tt hep-ph/9810299}.}

\lref\Tornqvdr{N.~Tornqvist and M.~Roos,
``Reply to Comment on 'Confirmation of the $\sigma$ meson',"
Phys. Rev. Lett. {\bf 78}, 1604 (1997).}

\lref\Scadroc{M.D.~Scadron,
``Comments on compositeness in the $SU(2)$ linear $\sigma$ model,"
Phys. Rev. {\bf D57}, 5307 (1998)
{\tt hep-ph/9712425}.}

\lref\Ishizg{M.~Ishida, S.~Ishida and T.~Ishida,
``Relation between scattering and production amplitude: case of intermediate $\sigma$ particle in $\pi \pi$ system,"
{\tt hep-ph/9712233}.}

\lref\Takam{K.~Takamatsu {\it et al.}, ``$\sigma$ particle in production processes," {\tt hep-ph/9712232}.}

\lref\Torr{N.A.~Tornqvist, ``The linear $U(3) \times U(3)$ $\sigma$--model, the $\sigma$(500) and the spontaneous breaking of symmetries," {\tt hep-ph/9711483}.}

\lref\Sved{M.~Svec, ``Mass and width of $\sigma$(750) scalar meson from measurements of $\pi^- p \to\pi^- \pi^+ n$ on polarized target,"
{\tt hep-ph/9711376}.}

\lref\Oller{J.A.~Oller and E.~Oset, ``Chiral symmetry amplitudes in the S wave isoscalar and isovector channels and the $\sigma$, $f^{(0)}(980)$, $a^{(0)}(980)$ scalar mesons,"
{\tt hep-ph/9710554}.}

\lref\Bramon{A.~Bramon, ~Riazuddin and M.D.~Scadron,
``Double counting ambiguities in the linear $\sigma$ model,"
J. Phys. G {\bf G24}, 1 (1997)
{\tt hep-ph/9709274}.}

\lref\Babouk{L.R.~Baboukhadia, V.~Elias and M.D.~Scadron,
``Linear $\sigma$ model linkage with nonperturbative QCD,"
J. Phys. {\bf G23}, 1065 (1997)
{\tt hep-ph/9708431}.}

\lref\Ishidz{S.~Ishida {\it et al.}, ``Further analysis on $\sigma$ particle properties," Prog. Theor. Phys. {\bf 98}, 1005 (1997).}

\lref\Malbo{A.P.~Malbouisson, M.B.~Silva-Neto and N.F.~Svaiter,
``One loop dimensional reduction of the linear $\sigma$ model,"
Physica {\bf A250}, 423 (1998)
{\tt hep-th/9705070}.}

\lref\Efrosin{V.P.~Efrosinin and D.A.~Zaikin,
``On certain consequences of the $SU(2)$ linear $\sigma$ model,"
Bull. Lebedev Phys. Inst. {\bf }, 13 (1990).}

\lref\Sander{M.~Sander and H.V.~von Geramb,
``$\pi \pi$, $K \pi$ and $\pi N$ potential scattering and a prediction of a narrow $\sigma$ meson resonance,"
Phys. Rev. {\bf C56}, 1218 (1997)
{\tt nucl-th/9703031}.}

\lref\Haradk{M.~Harada, ``Existence of $\sigma$ meson in $\pi \pi$ scattering," {\tt hep-ph/9702332}.}

\lref\daRocha{C.A.~da Rocha, G.~Krein and L.~Wilets,
``Ghost poles in the nucleon propagator in the linear $\sigma$ model approach and its role in $\pi N$ low--energy theorems,"
Nucl. Phys. {\bf A616}, 625 (1997) 
{\tt nucl-th/9701017}.}

\lref\Chanowitz{M.S.~Chanowitz, ``Tree unitary $\sigma$ models and their application to strong $W W$ scattering," {\tt hep-ph/9612240}.}

\lref\Aouisse{Z.~Aouissat, P.~Schuck and J.~Wambach,
``The linear $\sigma$ model in the $1/N$ expansion via dynamical boson mappings and applications to $\pi \pi$ scattering,"
Nucl. Phys. {\bf A618}, 402 (1996)
{\tt nucl-th/9611048}.}

\lref\Tornqc{N.A.~Tornqvist and M.~Roos,
``Reply to the comment by Harada, Sannino and Schechter on 'confirmation of
 the $\sigma$ meson'," Phys. Rev. Lett. {\bf 77}, 2333 (1996)
{\tt hep-ph/9610527}.}

\lref\Ishidp{M.~Ishida, ``Property of $\sigma$(600) and chiral symmetry," Prog. Theor. Phys. {\bf 96}, 853 (1996) {\tt hep-ph/9610326}.}

\lref\Harada{M.~Harada, F.~Sannino and J.~Schechter,
``Comment on 'Confirmation of the $\sigma$ meson',"
Phys. Rev. Lett. {\bf 78}, 1603 (1997) {\tt hep-ph/9609428}.}

\lref\Sonoda{H.~Sonoda, ``Note on the off-shell equivalence between the linear and nonlinear $\sigma$  models," Nucl. Phys. {\bf B490}, 457 (1997) {\tt hep-th/9609132}.}

\lref\Tornb{N.A.~Tornqvist,``The scalar $q\bar q$ nonet and confirmation of the broad $\sigma$ approximately (500) meson," 
{\tt hep-ph/9608464}.}

\lref\Flego{S.P.~Flego, A.H.~Blin, B.~Hiller and M.C.~Nemes,
``Pion, sigma and nucleon propagators in the linear $\sigma$ model,"
Z. Phys. {\bf A352}, 197 (1995).}

\lref\Isgur{N.~Isgur and J.~Speth, ``Comment on 'Confirmation of the sigma meson'," Phys. Rev. Lett. {\bf 77}, 2332 (1996).}

\lref\Ishida{S.~Ishida {\it et al.}, ``An analysis of $\pi \pi$ scattering phase shift and existence of $\sigma (555) $ particle," Prog. Theor. Phys. {\bf 95}, 745 (1996) {\tt hep-ph/9610325}.}

\lref\Torn{N.A.~Tornqvist and M.~Roos, ``Resurrection of the sigma meson," Phys. Rev. Lett. {\bf 76}, 1575 (1996) {\tt hep-ph/9511210}.}

\lref\Svec{M.~Svec, ``Study of $\sigma (750)$ and $\rho^0 (770)$ production in measurements of $\pi N$ (polarized) $\to \pi^+ \pi^- N$ on a polarized target at 5.98 GeV/c, 11.85 GeV/c and 17.2 GeV/c,"
Phys. Rev. {\bf D53}, 2343 (1996) {\tt hep-ph/9511205}.}

\lref\GasserBern{J. Gasser, ``$\pi\pi$ scattering at low energy: status report'', in {\sl Proceedings, Dubna May 1998 Positronium}, {\tt hep-ph/9809280}.}    

 \lref\Ugm{U.G. Meisssner, ``Chiral dynamics: where are the scalars?'', Boston University preprint BUTP-90/36, KEK library received 01-17-91.}

\lref\scadbla{M.D. Scadron,  ``Comments on compositeness in the $SU(2)$ linear $\sigma$ model'', {\tt hep-ph/9510344}.}

\newsec{Introduction}

One of the outstanding problems in high energy physics is the quantitative description of strong interactions at low energies. The systematic procedure known as chiral perturbation theory\ref\gsle{J. Gasser, H. Leutwyler, Ann. Phys. (NY) 158(1984) 142; Nucl. Phys. B250 (1985) 465.}, exploiting the global symmetries of that regime, expands the effective lagrangian in powers of derivatives and assigns to each term a coupling constant to be fit from experiment. Provided we focus on processes at low enough energies, chiral perturbation theory is thus supposed to yield good agreement with data \ref\holstein{J.F. Donoghue, E. Golowich, B.R. Holstein, {\sl Dynamics of the Standard Model} (1992) Cambridge University Press.}. Unfortunately, chiral perturbation theory does not give very good results on the scattering lengths of pion-pion scattering, which are very relevant experimental quantities in the limit of zero momentum, that is to say, where chiral perturbation theory should work best. 

Also, there is mounting evidence \Svec,  \Torn,  \Tornb, \Isgur, \Tornqvdr, \Harada, \Tornqc,   \Ishida, \Ishidp,  \Ishizg, \Sved, \Takam, \Haradk, \Sander,  \Oller, \Isck, \Lucio,    though still somewhat controversial, that a wide scalar resonance in the vicinity of 600 MeV exists. This resonance can be identified naturally with the $\sigma$ particle of the original {\sl linear} $\sigma$--model; this idea has been pursued in various forms recently \Efrosin, \scadbla, \Scadroc, \Flego, \Sonoda, \Aouisse, \daRocha, \Babouk, \Bramon, \Torr,   \Ishidz, \Malbo, \Tornqvism, \Napsuc.

With the above two motivations, we wish to explore the phenomenological consequences of the linear $\sigma$--model in $\pi\pi\to\pi\pi$ scattering at very low energies, and check that it gives a better fit to the experimental data than chiral perturbation theory. The sole guiding principle will be chiral symmetry, whose Ward identities we will exploit to modify the various vertices to take into account the large width of the $\sigma$ resonance. 

It is pleasant that the simple linear model works better than chiral perturbation theory, although it certainly cannot be the whole story, since, as we shall see, unitarity checks demonstrate that for relatively small momenta more resonances should be taken into account.  Let us stress that chiral perturbation theory at higher loops does get better at the disgraceful price of any non-renormalizable field theory, namely a growing number of couplings or fit parameters. 

The paper is organized as follows. First, we recall the naive linear 
$\sigma$--model, including the explicit chiral symmetry breaking term to account for $m_\pi\ne0$. The crucial point is that the breaking is soft, so that when we   include the width $\Gamma_\sigma$ of the $\sigma$ in its propagator, we can exploit the chiral Ward identities to modify accordingly the vertices. The chiral Ward identities are still satisfied by the final lagrangian (with parameters $m_\pi$, $f_\pi$, $m_\sigma$ and $\Gamma_\sigma$), from which we compute the amplitudes in the various isospin and angular momentum channels of experimental relevance. We then use the expression for $\Gamma_\sigma$ from the decay $\sigma\to\pi\pi$ to perform a simple and succesful one--parameter ($m_\sigma$) fit to experimental data. As an important check, we  analyze the various unitarity conditions on these amplitudes, then we discuss our results, and conclude. 

\newsec{The lagrangian and chiral Ward identities}
The guiding principle throughout is chiral symmetry. To realize it linearly,  we consider a complex spin zero ({\bf 2},{\bf 2}) field $D(x)$ with, of course, a Mexican hat potential of the form $\tilde\lambda(D^\dagger D -f_\pi^2)^2$. After field redefinitions around the true vacuum $\vev{D}\ne0$, the massless Goldstone bosons of chiral symmetry form a pseudoscalar isotriplet $\vec\pi(x)$, and the left--over massive field is a scalar isosinglet $\sigma(x)$, akin to the Higgs field of the standard model:
\eqn\lsm
{\eqalign{ {\cal L}&\ =  \12\partial_\mu\vec\pi \cdot \partial^\mu\vec\pi +  \12\partial_\mu\sigma \partial^\mu\sigma - \12 m_\sigma^2  \sigma^2
 - {{\tilde\lambda} f_\pi}  (\vec\pi\cdot\vec\pi  +\sigma^2)\sigma  + {{\tilde\lambda} \over 4} \left( \vec\pi\cdot\vec\pi + \sigma^2\right)^2   \cr}}
with \eqn\lamba{ \tilde{\lambda} = -{m_\sigma^2\over 2 f_\pi^2} }
So far, the chiral symmetry is spontaneously broken and the chiral Ward identities are exact. 
Chiral symmetry is explicitly broken by  the linear term  
\eqn\deltal{ \Delta{\cal L} =  a \sigma, }
which induces a non-vanishing pion mass $m_\pi^2 = -a/f_\pi$. Furthermore, the cubic and quartic couplings in 
the lagrangian \lsm\ undergo the modification \eqn\mod{ \tilde\lambda \to \lambda = -{m_\sigma^2 -m_\pi^2\over 2 f_\pi^2}} 

So far, this is all very well known. We claim that it is important to take into account consistently the fact that the field $\sigma$ is (very) unstable. For instance, if $m_\sigma>2m_\pi$, its tree--level width is 
\eqn\wid{ \Gamma\left(\sigma\to \pi\pi\right)=
{3m_\sigma^3 \over32\pi f_\pi^2} (1-\varepsilon)^2\sqrt{1-4\epsilon} }
where we have introduced the convenient shorthand $\varepsilon=(m_\pi/m_\sigma)^2$. For the sake of generality, we will treat $\Gamma_\sigma$ as a free parameter, referring to its tree--level value $\Gamma_\sigma(m_\sigma)$ in \wid\ explicitly only as need arises. In strict analogy with the Higgs field in the standard model, the $\sigma$ width $\Gamma_\sigma$ grows very fast with its mass\foot{Thus, for instance, $\Gamma_\sigma(350 )= 65$, $\Gamma_\sigma(500 )= 310$, $\Gamma_\sigma(650 )= 785$, all in MeV.}. The effect of the width of the $\sigma$ field is to modify its propagator from the usual $i\left( q^2-m_\sigma^2 \right)^{-1}$ to
\eqn\prop{ \Delta_\sigma(q)= i\left( q^2-m_\sigma^2 +i\Gamma_\sigma m_\sigma \, \theta( q^2-4m_\pi^2) \right)^{-1},  }
where the step function ensures that the imaginary piece in the denominator appears only when the momentum of the propagator is  above the kinematical threshold for $\sigma$ decay. 

Thus, in the physical process of $\pi\pi\to\pi\pi$ scattering, which we shall consider shortly, the propagator of the $\sigma$  picks up the correction due to the width only in the $s$--channel, not in the $u$-- nor the $t$--channels.

The crucial point is that, in the linear $\sigma$ model, chiral symmetry is responsible for important cancellations which imply, notably, that the pion coupling is always derivative in the limit of soft pion momenta. Once we include the large width $\Gamma_\sigma$, this cancellation is brutally spoiled. Equivalently, the chiral Ward identities are violated by the inclusion of the width in the $\sigma$ propagator. 
Enforcing, however, the chiral Ward identities on the vertices of the lagrangian  implies that the latter pick up modifications related to the width $\Gamma_\sigma$. These vertex corrections depend on the kinematical variables (the incoming momenta) in a particular way, dictated by chiral symmetry. For instance, the $\sigma \pi^i\pi^j$ Feynman rule reads now \eqn\fey{ V_{\sigma \pi^i\pi^j}= {-i\over f_\pi} \delta^{ij} \left( m_\sigma^2 -m_\pi^2 -i\Gamma_\sigma m_\sigma \theta(q^2-4m_\pi^2)\right)}
where $q^\mu$ is the momentum of the $\sigma$.

 The modification to the $\pi^4$ contact term is quite non--trivial and somewhat cumbersome to write out: the shift $m_\sigma^2\to m_\sigma^2 -i\Gamma_\sigma m_\sigma$ takes place depending on the external momenta. This dependence can be spelled out in terms of the momenta of the four pions entering the vertex:
\eqn\cuatropi{ V_{\pi^i\pi^j\pi^k\pi^\ell}= -2i \left\{ 
\left( \lambda + c_s\right) \delta^{ij}\delta^{k\ell} +\left( \lambda + c_t\right) \delta^{ik}\delta^{j\ell} +\left( \lambda + c_u\right) \delta^{i\ell}\delta^{jk} \right\}  }
where (the incoming momentum  of a pion of isospin $i$ is $p_i$)
\eqn\ces{ \eqalign{ & c_s = {i\Gamma_\sigma m_\sigma \over 2 f_\pi^2 } \theta \left( (p_i+p_j)^2-4m_\pi^2\right) \cr
 & c_t = {i\Gamma_\sigma m_\sigma \over 2 f_\pi^2 } \theta \left( (p_i+p_k)^2-4m_\pi^2 \right) \cr
 & c_u = {i\Gamma_\sigma m_\sigma \over 2 f_\pi^2 } \theta \left( (p_i+p_\ell)^2-4m_\pi^2\right) \cr }}
and we recall equation \mod\ whereby $\lambda = -\left( m_\sigma^2- m_\pi^2 \right)/( 2 f_\pi^2)$.

The procedure sketched above to incorporate the width, consistently with the chiral Ward identities, is very powerful. Similar considerations have been set forth recently in the analogous problem of incorporating the width of the electroweak gauge bosons to the fermionic vertices of the standard model \ref\gabi{G. L\'opez--Castro, J.L. Lucio, J. Pestieau, Mod. Phys. Lett. A6 (1991) 3679}, \ref\papav{J. Papavassiliou, A. Pilaftsis, Phys. Rev. Lett. 75 (1995) 3660.}, \ref\tropa{E.N. Argyres et al., Phys Lett. B358 (1995) 339.}.

Let us focus on the chiral Ward identities which force us  to modify the Feynman rules derived from the lagrangian when  the $\sigma$ propagator changes to \prop. They can be derived with the usual method of gauging the chiral symmetry transformation and then setting the chiral gauge boson to zero. Consider, for instance, the following:
\eqn\quira{ V_{\pi^i\pi^j\sigma\sigma} = V_{\sigma\sigma \sigma} \Delta_\sigma(p_j)   V_{\sigma\pi ^i\pi^j} } Note that $p_j$ is the momentum of a pion, so that $p_j^2 =m_\pi^2$ if it is on--shell. This equation {\sl defines} the $\pi\pi\sigma\sigma$ vertex. Similarly, the chiral Ward identity satisfied by the $\pi^4$ Feynman rule is 
\eqn\quirb{ V_{\pi^i\pi^j\pi^k\pi^\ell} = V_{\pi^k\pi^\ell \sigma} \Delta_\sigma(p_j)   V_{\sigma\pi ^i\pi^j} +V_{\pi^i\pi^k \sigma} \Delta_\sigma(p_k)   V_{\sigma\pi ^j\pi^\ell} +V_{\pi^i\pi^\ell \sigma} \Delta_\sigma(p_\ell)   V_{\sigma\pi ^j\pi^k} }
Obviously, these relations hold at tree level before chiral symmetry breaking, that is to say, when $m_\pi=0$, and also $\Gamma_\sigma=0$. Powefully, they also hold when $m_\pi\not=0$ and/or when $\Gamma_\sigma\ne0$, to all orders in perturbation theory. This can be proved easily using the enormous advantage that the linear sigma model is a well--defined (renormalizable) field theory. The chiral Ward identities \quira\ and \quirb\ are satisfied by \prop--\cuatropi, of course.

Since the vertex modifications ensure the preservation of exact chiral Ward identities, they  also guarantee, for instance, that the pion couplings remain derivative as they should. 

\newsec{Pion scattering revisited}

As an illustration of the power of this implementation of chiral symmetry, we proceed to evaluate, at tree level, the amplitude for $\pi\pi$ scattering. Clearly, we do not expect the result to be the perfect answer, since the only resonance we will take into account is the $\sigma$. In particular, we shall see that not taking into account the vector meson $\vec{\rho}^\mu$ is a rather bad approximation in the $I=1$, $\ell=1$ amplitude. Nevertheless, our results are in better agreement with experimental data than those of chiral perturbation theory. Let us emphasize that the kinematical region where we compare both predictions, namely at very low momenta, is precisely where chiral perturbation theory should be exact. This lends further support to the real existence 
of $\sigma$ as a strong resonance. 

At tree level, four diagrams contribute to $\pi\pi\to\pi\pi$:  
the four--pion contact term, and the exchange of a $\sigma$ in the three $s$, $t$ and $u$ channels. The transition matrix 
 is of the form $T_{ij}^{k\ell}$, where  the indices $i$ and  $j$
 (respectively, $k$ and $\ell$) label the isospin of the initial (respectively, final) pions. We use  projectors $\Pi^{(I)}\big|_{ij}^{k\ell}$ in each of the isospin channels ($I=0$, 1 and 2) and then decompose in partial waves ($p$ is the modulus of the three-momentum of any of the four pions in the center of mass) following the traditional normalization
\eqn\til{ T^{(I)}_\ell (p) = {1\over64\pi} \int_{-1}^1 {\rm d}x\; T^{(I)}(p,x) P_\ell(x) } whereby, in terms of the partial wave shifts $\delta^{(I)}_\ell$, we have
\eqn\partwav{ T^{(I)}_\ell = \sqrt{s\over s-4 m_\pi^2} {\rm e}^{i\delta^{(I)}_\ell} \; {\rm sin} \delta_\ell^{(I)} } with $s=4(p^2+m_\pi^2)$. We shall return to these below when we study unitarity.
Furthermore,  results of analyses of experimental data are often quoted in terms of $a^{(I)}_\ell$ and $b^{(I)}_\ell$, which are just the first two coefficients in the Taylor expansion of the real part of $T^{(I)}_\ell(p) $ around $p=0$:
\eqn\expan{ {\rm Re}\;  T^{(I)}_\ell (p) = \left(p^2 \over m^2_\pi \right)^\ell \left(  a^{(I)}_\ell  +{ p^2 \over m^2_\pi }b^{(I)}_\ell +\ldots \right)}

Due to the structure of the Feynman rules dictated by chiral Ward identities, the width $\Gamma_\sigma$ contributes, in the Born approximation, only to $T^{(0)}_0$. Also, $T^{(0)}_\ell (p)$ and $T^{(2)}_\ell (p)$ vanish for odd $\ell$, whereas $T^{(1)}_\ell (p)$ is zero for even $\ell$. This is consistent with general symmetry arguments. In our computation, at this level of simplicity, we find also \eqn\igual{ T^{(2)}_\ell (p) =T^{(0)}_\ell (p) \qquad \forall \ell\ne0 }
 
The experimental knowledge of pion scattering near threshold is rather poor. The relatively badly measured scattering lengths and 
ranges are \holstein\  $a_0^{(0)}$, $b_0^{(0)}$, $a_0^{(2)}$, $b_0^{(2)}$, $a_1^{(1)}$, $a_2^{(0)}$ and  $a_2^{(2)}$, These seven numbers should come out of our computation with only $m_\sigma $ and $\Gamma_\sigma$ as free parameters (actually, only the first two of these quantitites, the s-wave isoscalar moments, depend on the width).  We shall consider  these seven data points  as uncorrelated, independent measurements with gaussian errors.
 
Explicitly, recalling that $\varepsilon=(m_\pi/m_\sigma)^2$ and letting also $\gamma=(\Gamma_\sigma/m_\sigma)^2$ as a free parameter, we find
\eqn\asbs{\eqalign{
&a^{(0)}_0= {1\over 32\pi}\;\left( m_\pi \over f_\pi\right)^2
 {7 - 27 \varepsilon - 12 \varepsilon^2  + 32 \varepsilon^3  + 7 \gamma
 + 2 \varepsilon \gamma \over 1 - 8 \varepsilon + 16 \varepsilon^2  + \gamma}   \cr
 &b^{(0)}_0= {1\over 8\pi}\;\left( m_\pi \over f_\pi\right)^2
 \Bigl[ 2 (1 - 4 \varepsilon)^2  (1 - \varepsilon)^2  (1 + 4 \varepsilon - 8 \varepsilon^2 )
 + \gamma (4- 10 \varepsilon - 69 \varepsilon^2  
 + 80 \varepsilon^3   - 32 \varepsilon^4  ) \cr  
 &  \qquad\qquad\qquad \qquad\qquad +  \gamma^2 (2  + 2 \varepsilon  - \varepsilon^2)\Bigr]   /  \left(1 - 8 \varepsilon + 16 \varepsilon^2  + \gamma\right)^2 \cr   
&a^{(0)}_2=a^{(2)}_2= {1\over 30\pi}\;\left( m_\pi \over f_\pi\right)^2 \; \varepsilon (1-\varepsilon)^2\cr                                          
 &b^{(0)}_2=b^{(2)}_2= {-1\over 5\pi}\;\left( m_\pi \over f_\pi\right)^2 \; \varepsilon^2 (1-\varepsilon)^2\cr 
&a^{(1)}_1= {1\over 24\pi}\;\left( m_\pi \over f_\pi\right)^2 (1-\varepsilon)^2\cr 
&b^{(1)}_1= {-1\over 6\pi}\;\left( m_\pi \over f_\pi\right)^2 \; \varepsilon (1-\varepsilon)^2\cr
&a^{(2)}_0= {-1\over 16\pi}\;\left( m_\pi \over f_\pi\right)^2 (1-\varepsilon)\cr 
&b^{(2)}_0= {-1\over 8\pi}\;\left( m_\pi \over f_\pi\right)^2 (1-\varepsilon)^2\cr   }}       

When we use \wid, $\gamma$ disappears and becomes a function of the single parameter $m_\sigma$. Note that the nonlinear limit $\varepsilon\to0$ of \asbs\ is {\sl independent} of $\gamma$, so that regardless of whether $\Gamma_\sigma$ remains finite, or, more physically, of how it blows up as $m_\sigma$ does, the tree-level results of chiral perturbation theory are easily recovered by shooting away the $\sigma$ resonance from the universe.

 We take as  experimental averages\holstein\  $a_0^{(0)}=.26\pm.05$, $b_0^{(0)}=.25\pm.03$, $a_0^{(2)}-.028\pm.012$, $b_0^{(2)}-.082\pm.008$, $a_1^{(1)}=.038\pm.002$, $a_2^{(0)}=.0017\pm.0003$ and  $a_2^{(2)}=.00013\pm.00030$.      
An overall fit to these seven numbers gives  $m_\sigma=700 {+ 800 \atop -150}$MeV, where the errors are determined by an increase in one of $\chi^2/dof$ over the value at the minimum. Clearly, the   $\chi^2$  distribution is very flat towards increasing values of $m_\sigma$;     $m_\sigma\ge$550 Mev is the only useful information. 

Of the seven numbers, if we eliminate the worst one ($a_1^{(1)}$ (presumably under strong influence from $\rho$ exchange, which we ignore), then the fit improves and it yields $m_\sigma= 590 { +220 \atop -90}$MeV. For completeness, note that the fit to only the scalar isoscalar values gives  
$m_\sigma= 525 { +85 \atop -45}$MeV. 

Overall, one may conclude that the data are, on the whole, consistent with a linear sigma resonance provided its mass is around 600 MeV (and thus its width also around 600 MeV). The errors on these numbers, from the pion data available, are substantial. 

Although the low--energy moments $a_\ell^{(I)}$ and 
$b_\ell^{(I)}$ are the relevant quantities for us, what is actually measured is a momentum--dependent phase shift, which can be split more or less in various isospin and angular momentum channels, $\delta=\sum_{I,\ell} \delta^{(I)}_\ell$, with
\eqn\fase{\delta^{(I)}_\ell = \12{\rm Arcsin} \left( 1 \sqrt{1-{4 m_\pi^2\over s}} \,{\rm Re} T^{(I)}_\ell[s,m_\sigma,\Gamma_\sigma] \right) }
From the analysis of the data available (five points  at $(\delta,\sqrt{s}/{\rm MeV})$=$(.07\pm.13,289),$ $(.21\pm .07,303),$ $(.13\pm .05,317)$, $(.20\pm.04,335)$ and $(.27\pm.04,367)$) in the $I=0$ and $\ell=0$ channel, we   fit    $m_\sigma=550 {+450 \atop -80}$MeV. Again   the error on the heavy side is huge: the $\chi^2$ distribution is very flat with increasing  $m_\sigma$.

\newsec{Unitarity}

Massaging expression \partwav, one gets unitarity constraints from the fact that $\delta_\ell^{(I)}$ are real.
 Weak constraints  are \eqn\uwa{ {s-4m_ \pi^2 \over s} \left|T_\ell^{(I)}\right|^2  <1 }
and
 \eqn\uwb{0 < \sqrt{s-4m_\pi^2\over s} \;{\rm Im}\, T_\ell^{(I)} < 1}
 A rather strong condition is  
\eqn\unis{\sqrt{ s-4m_\pi^2 \over s} \left| {\rm Re}\, T_\ell^{(I)} \right|\le \12 }
whereas exact unitarity is achieved iff
\eqn\exun{ {\rm Im}\, T_\ell^{(I)} = \sqrt{ s-4m_\pi^2 \over s}  \left| {\rm Re}\, T_\ell^{(I)} \right|^2 }
from which the optical theorem can be derived. Clearly, all these constraints depend on the energy scale. Since there are many other resonances in nature heavier than the $\sigma$, we should not worry much about possible unitarity violations at high momenta (say, above 1 GeV); still, we should make sure that the behaviour near threshold is not pathological: unitarity should be  well preserved at center of mass momenta lower than, say, the pion mass.

It turns out that there is no problem with unitarity at center of mass   momenta lower than the 400 MeV. Unfortunately, unitarity does not constrain $m_\sigma$ from above in any meaningful way.

\newsec{Conclusions}

We have enhanced the linear sigma model by enforcing chiral Ward identities which take into account the (large) sigma width. 
We have found that low energy pion scattering data supports the existence of a wide $\sigma$ field with mass around 600 MeV (actually $m_\sigma= 590 { +220 \atop -90}$MeV), provided we exclude the datum in the vector isovector channel. The advantage of keeping the $\sigma$ as a true resonance in the effective low energy theory of strong interactions is not only that its inclusion simulates more or less the results of chiral perturbation theory to one loop, as emphasized by skeptics long ago \Ugm, but also, more crucially, that   this opens the door to more industrious analyses of the whole scalar spectrum, including glueballs. 

\bigbreak\bigskip\bigskip\centerline{{\bf Acknowledgements}}\nobreak
This work was supported in part by CONACYT through projects 3979P-E9608, 25504-E,  and C\'atedra Patrimonial II de Apoyo a los Estados, and by DGAPA--UNAM through IN103997.


\listrefs
\bye